\documentclass[12pt]{article}
\usepackage{geometry}
\usepackage{amsmath}
\usepackage{amssymb,epsfig,subfigure}
\usepackage{graphicx}
\usepackage{cite}
\numberwithin{equation}{section}

\newcommand{\be}{\begin{equation}}
\newcommand{\ee}{\end{equation}}
\newcommand{\ba}{\begin{aligned}}
\newcommand{\ea}{\end{aligned}}

\def\m1{\left(-1\right)^{F_i}}

\makeatletter
\def\sla@#1#2#3#4#5{{%
  \setbox\z@\hbox{$\m@th#4#5$}%
  \setbox\tw@\hbox{$\m@th#4#1$}%
  \dimen4\wd\ifdim\wd\z@<\wd\tw@\tw@\else\z@\fi
  \dimen@\ht\tw@
  \advance\dimen@-\dp\tw@
  \advance\dimen@-\ht\z@
  \advance\dimen@\dp\z@
  \divide\dimen@\tw@
  \advance\dimen@-#3\ht\tw@
  \advance\dimen@-#3\dp\tw@
  \dimen@ii#2\wd\z@  \raise-\dimen@\hbox to\dimen4{%
    \hss\kern\dimen@ii\box\tw@\kern-\dimen@ii\hss}%
  \llap{\hbox to\dimen4{\hss\box\z@\hss}}}}
\def\slashed#1{%
  \expandafter\ifx\csname sla@\string#1\endcsname\relax
    {\mathpalette{\sla@/00}{#1}}%
  \else
    \csname sla@\string#1\endcsname
  \fi}
\makeatother

\begin{document}


\thispagestyle{empty}
\begin{flushright}\footnotesize
\texttt{CALT-68-2698}\\
\vspace{2.1cm}
\end{flushright}

\renewcommand{\thefootnote}{\fnsymbol{footnote}}
\setcounter{footnote}{0}

\begin{center}
{\Large\textbf{\mathversion{bold} An Instanton Toolbox for F-Theory Model Building}\par}

\vspace{2.1cm}

\textrm{Joseph Marsano, Natalia Saulina and
Sakura Sch\"afer-Nameki}

\vspace{1cm}

\textit{California Institute of Technology\\
1200 E California Blvd., Pasadena, CA 91125, USA } \\
\texttt{marsano, saulina, ss299 @theory.caltech.edu}

\bigskip


\par\vspace{1cm}

\textbf{Abstract}\vspace{5mm}
\end{center}

\noindent
Several dimensionful parameters needed for  model building can be engineered in a certain class of $SU(5)$ F-theory GUTs by adding extra singlet fields which are localized along pairwise intersections of D7-branes.  The values of these parameters, however, depend on dynamics external to the GUT which causes the singlets to acquire suitable masses or expectation values.
In this note, we demonstrate that D3-instantons which wrap one of the intersecting D7's can provide precisely the needed dynamics to generate several important scales, including the supersymmetry-breaking scale and the right-handed neutrino mass.  Furthermore, these instantons seem unable to directly generate the $\mu$ term suggesting that, at least in this class of models, it should perhaps be tied to one of the other scales in the problem.
More specifically, we study the simple system consisting of a pair of D7-branes wrapping del Pezzo surfaces which intersect along a curve $\Sigma$ of genus 0 or 1 and classify all instanton configurations which can potentially contribute to the superpotential.  This allows one to formulate topological conditions which must be imposed on $\Sigma$ for various model-building applications.
Along the way, we also observe that the construction of \cite{Heckman:2008es} which engineers a linear superpotential in fact realizes an O'Raifeartaigh model at the KK scale whose 1-loop Coleman-Weinberg potential generically leads to a metastable, long-lived SUSY-breaking vacuum.

\vspace*{\fill}

\setcounter{page}{1}
\renewcommand{\thefootnote}{\arabic{footnote}}
\setcounter{footnote}{0}

 \newpage


\tableofcontents


\section{Introduction}

F-theory compactifications on local Calabi-Yau four-folds appear to be a very promising arena for engineering realistic supersymmetric Grand Unified Theories (GUTs) in string theory \cite{Donagi:2008ca,Beasley:2008dc,Beasley:2008kw}. The properties of these models have been further explored in \cite{Hayashi:2008ba,Aparicio:2008wh,Buchbinder:2008at,Heckman:2008es,Marsano:2008jq}.  In such constructions, the gauge group is housed on a stack of 7-branes wrapping a 4-cycle $S_{GUT}$ and charged matter is localized on curves where this stack is intersected by additional 7-branes.  Both the spectrum and structure of the superpotential can be determined completely by the topology of the four-fold near $S_{GUT}$ and any nontrivial gauge bundles present there.

Supersymmetric GUTs exhibit a number of dimensionful parameters with scales typically lower than $M_{GUT}$ whose origin must be explained.  These include the supersymmetry-breaking scale and the supersymmetric Higgs mass term, $\mu$.  In addition, models which implement the seesaw mechanism in the neutrino sector must also introduce a mass for the right-handed neutrino.  Quite nicely, it was observed in \cite{Beasley:2008kw} that all of these parameters can be naturally incorporated into one class of $SU(5)$ GUTs in \cite{Beasley:2008dc,Beasley:2008kw} using a very simple configuration, namely a pair of intersecting D7-branes which meet the GUT stack at a point of triple intersection.  This setup is depicted in figure \ref{fig:intbranes}.

To see how such a simple system can be so versatile, recall that open strings stretching from one of the D7-branes to the GUT stack engineer charged matter in the $\mathbf{5}$ and $\mathbf{\overline{5}}$ representations.  Bifundamentals between the D7-branes themselves engineer a GUT singlet $\mathbf{1}$ and the Yukawa coupling which arises at the triple intersection point takes the form
\begin{equation}
\mathbf{5}\times\mathbf{\overline{5}}\times\mathbf{1} \,.
\label{oursup}\end{equation}
If we identify the $\mathbf{5}$ and $\mathbf{\overline{5}}$ as messenger fields $f$ and $\bar{f}$ for gauge mediation, then this is simply the superpotential for ordinary gauge mediation \cite{Beasley:2008kw,Marsano:2008jq}{\footnote{As usual, both $f$ and $\bar{f}$ are 4d ${\cal{N}}=1$ chiral superfields.}}
\begin{equation}
W_{OGM}\sim Xf\bar{f} \,,
\end{equation}
where the GUT singlet field is denoted by $X$.
On the other hand, if we identify the $\mathbf{5}$ as the $SU(5)$ multiplet containing the up-type Higgs, $H_u$, and the $\mathbf{\overline{5}}$ as the multiplet containing the down-type Higgs, $H_d$, then we instead write this superpotential as
\begin{equation}
W_{\mu}\sim XH\bar{H}\,.
\end{equation}
We see that this can give rise to a nonzero $\mu$ term provided some physics external to the GUT stack causes $X$ to obtain a nonzero bosonic expectation value \cite{Beasley:2008kw}.  Note that interactions of this type also appear in one of the two scenarios discussed in \cite{Marsano:2008jq} for coupling the Higgs and messenger sectors in models of gauge mediated supersymmetry-breaking.  In that case the $\mathbf{5}$ is identified as $H$ or $f$ and the $\mathbf{\overline{5}}$ as $\bar{f}$ or $\bar{H}$ in order to generate direct couplings between the Higgs and messenger sectors through the interactions
\begin{equation}W_{\rho}\sim X H \bar{f}\end{equation}
or the analogous one with $H\bar{f}$ replaced by $\bar{H}f$.  In this case, we can obtain the couplings
\begin{equation}W_{\rho}\sim \rho H\bar{f} + \tilde{\rho}\bar{H}f\label{rhoterms}\end{equation}
where $\rho$ and $\tilde{\rho}$ are bosonic expectation values of our GUT singlet fields.  Finally, if we identify the $\mathbf{5}$ as the $SU(5)$ multiplet containing $H_u$ and the $\mathbf{\overline{5}}$ as one of the $SU(5)$ matter multiplets, denoted $\Phi_{\mathbf{\overline{5}}}$, then we get the superpotential
\begin{equation}
W_{N_R}\sim X H \Phi_{\mathbf{\overline{5}}} \,.
\end{equation}
In this case, the GUT singlet field $X$ is playing the role of the right-handed neutrino, $N_R$.  This can lead to neutrino masses via the seesaw mechanism provided $X$ becomes massive \cite{Beasley:2008kw}.

In each of these situations, we desire some dynamical mechanism to give a mass or suitable expectation value to matter localized at the intersection of two D7-branes.  It has already been established that a supersymmetry-breaking $F$-component expectation value can be induced at a small scale by using D3-instantons to generate a Polonyi superpotential in this setup \cite{Heckman:2008es}{\footnote{Previous studies of nonperturbatively generated superpotentials in F-theory also include \cite{Witten:1996bn,Ganor:1996pe,Robbins:2004hx,Saulina:2005ve,Kallosh:2005gs,Lust:2005cu,Tsimpis:2007sx}.}}. An important condition, which has to be satisfied in order for the D3-instantons to contribute and which was discussed in v2 of \cite{Heckman:2008es}, is the trivializability of the world-volume fluxes on the D3-instanton. We will discuss this below. 

In this note,
we further demonstrate that D3-instantons can in fact trigger the expectation values and mass terms needed to generate all of the scales above except for the $\mu$ term, whose generation seems to be obstructed by an important feature of doublet-triplet splitting in this class of $SU(5)$ GUTs.
  The fact that $\mu$ is singled out seems to suggest that its generation in these models must be tied to one of the other scales at our disposal.  This is of course a welcome feature for models with low-scale supersymmetry as one expects the $\mu$ term to sit naturally near the soft mass scale.  One mechanism for accomplishing this in F-theory has been suggested in \cite{Marsano:2008jq} and is based on the phenomenological framework developed by Ibe and Kitano \cite{Ibe:2007km}.  A different approach to the generation of $\mu$ in F-theory GUTs is expected to appear very soon \cite{HV}.  Note also that using instantons to induce supersymmetry-breaking, neutrino masses and other dimensionful parameters is  not a new idea and has in fact been implemented in a variety of different contexts in recent years  \cite{Blumenhagen:2006xt,Ibanez:2006da,Buican:2006sn,Cvetic:2007ku,Argurio:2007qk,Antusch:2007jd,Aharony:2007pr,Aharony:2007db,Aganagic:2007py,Ibanez:2007tu,Cvetic:2008hi,Buican:2008qe,Cvetic:2007qj,Cvetic:2008mh}.

In the simple case of intersecting D7-branes that wrap del Pezzo surfaces, we are able to classify all supersymmetric D3-instanton configurations which can lead to nontrivial superpotential couplings.
  Which superpotential terms are actually generated is determined by the class of the curve $\Sigma$ along which the D7-branes intersect.  We then provide several explicit examples which can be used to introduce supersymmetry-breaking, neutrino masses, and $\rho$ couplings of the form \eqref{rhoterms}.  
Finally, in our discussions of supersymmetry-breaking, we observe that the Polonyi construction of \cite{Heckman:2008es} actually realizes an O'Raifeartaigh model of the type studied in \cite{Shih:2007av} at the Kaluza-Klein scale, whose 1-loop Coleman-Weinberg potential lifts the pseudo-moduli space to yield a long-lived, metastable supersymmetry-breaking vacuum.
Throughout, we restrict our attention to single-instanton contributions because we expect that these will generally dominate any multi-instanton ones\footnote{Multi-instanton effects in this context were studied in \cite{Heckman:2008es}.}.

The organization of this note is as follows.  In  section \ref{sec:setup} we first review some details of the general setup and summarize our classification of the supersymmetric instanton configurations that generate nontrivial superpotential couplings.  We also discuss constraints that arise when our system is coupled to an $SU(5)$ F-theory GUT and the trouble associated with trying to directly generate the $\mu$ parameter.  In the rest of the note, we provide several simple examples.  In section \ref{sec:linear} we study linear superpotentials, which can be used for supersymmetry-breaking or generation of the $\rho$ coupling \eqref{rhoterms}.  We also review the construction of \cite{Heckman:2008es} and discuss how this system engineers the O'Raifeartaigh model of \cite{Shih:2007av}.  We then turn in section \ref{sec:quadratic} to quadratic superpotentials, which can be used to generate neutrino masses or $\rho$ couplings.  In section \ref{sec:mixed} we study a "mixed" superpotential involving a pair of chiral fields with opposite $U(1)$ charges, which realizes a Fayet model of supersymmetry-breaking.  Finally, we describe the details of our instanton zero mode analysis in Appendix \ref{app:instantons}.


\section{Setup and Summary of Instanton Contributions}
\label{sec:setup}

\begin{figure}\begin{center}
\subfigure[Triple intersection where our D7-branes meet the GUT stack]
{\epsfig{file=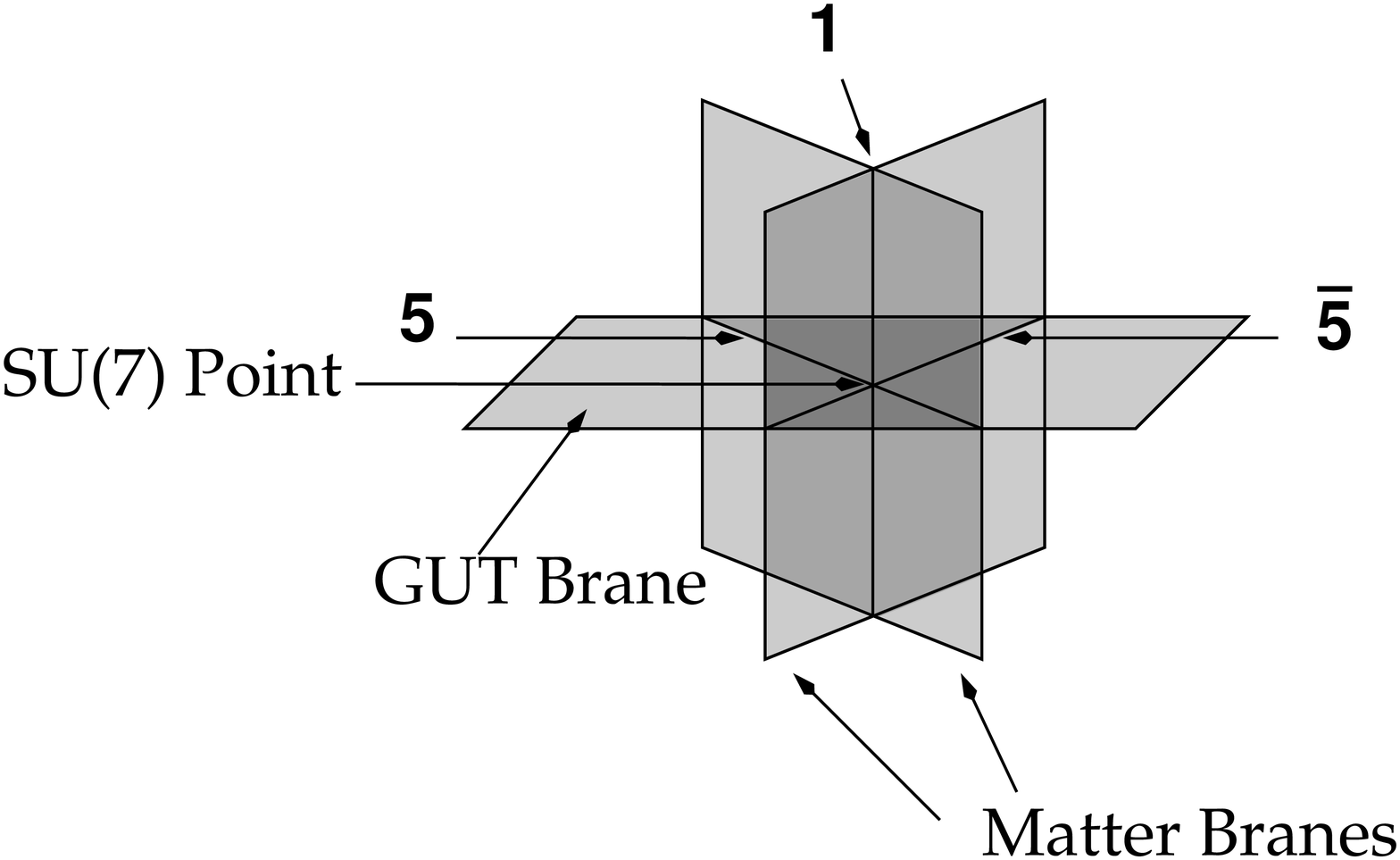,width=0.54\textwidth
}\label{fig:intbranes}}
\label{fig:Polonyi}
\hspace{0.1\textwidth}
\subfigure[General setup of intersecting D7-branes that we study in this note.]
{\epsfig{file=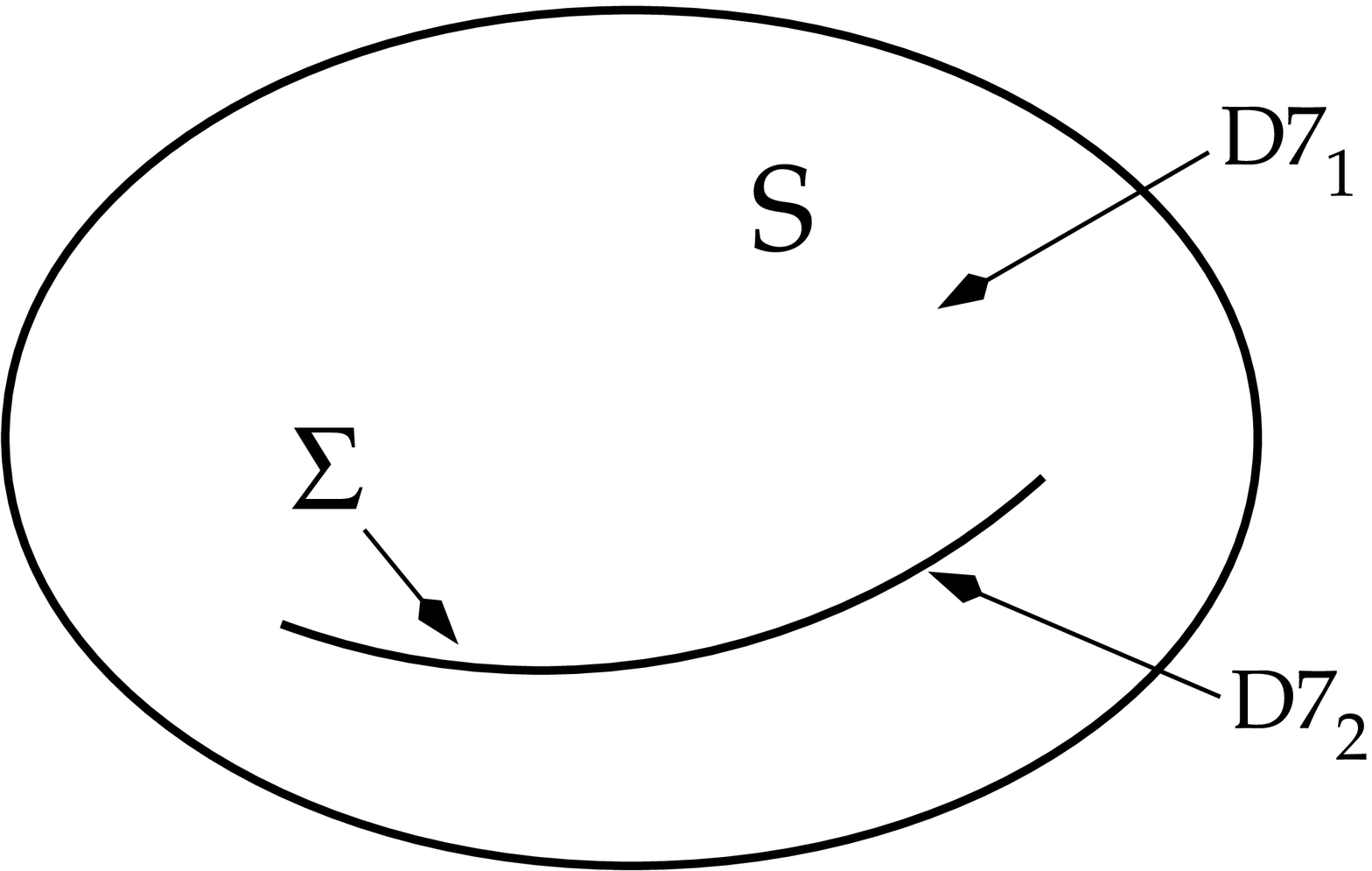,width=.33\textwidth
}\label{fig:Pol}}
\caption{Pair of intersecting D7-branes and their coupling to a 7-brane GUT}
\end{center}
\end{figure}

In this section we describe our basic setup, set notation, and present a general classification of D3-instanton configurations which can generate nontrivial superpotential terms.  The system under study is an extremely simple one, namely a pair of intersecting D7-branes.  For the purposes of bundle calculations, we will always assume that the D7's wrap del Pezzo surfaces.  This is motivated partly by their simplicity and partly by our interest in adhering to the local philosophy of \cite{Beasley:2008dc,Beasley:2008kw}, wherein one considers models for which a strict $M_{Pl}\rightarrow\infty$ limit can be taken at least in principle.

In the following, we shall be interested in studying the effects of single D3-instantons, which are coincident with one of the D7-branes.  As such, we single out one of the D7-branes, D7$_1$, and denote the 4-cycle that it wraps by $S$.  The second D7-brane, D7$_2$, then intersects $S$ along a curve $\Sigma$ while the D3-instanton whose effects we shall study wraps all of $S$.  This setup is depicted in figure \ref{fig:Pol}.

The D7-branes each house their own $U(1)$ gauge group, which we denote by $U(1)_i$ for D7$_i$.  We further obtain charged matter localized on $\Sigma$ and can induce a net chirality in the massless spectrum by turning on suitable supersymmetric $U(1)_i$ gauge bundles, $V_i$.  Throughout this note, we will typically use the notation $X$ or $X_I$ for chiral superfields with charges $(+,-)$ under $U(1)_1\times U(1)_2$ and $\tilde{X}$ or $\tilde{X}_I$ for chiral superfields with charges $(-,+)$.  The numbers $n_X$ and $n_{\tilde{X}}$ of massless chiral multiplets of type $X$ and $\tilde{X}$ can be computed, for instance, along the lines of \cite{Beasley:2008dc} as
\begin{equation}\begin{split}
n_{X}&=h^0(\Sigma,K_{\Sigma}^{1/2}\otimes V_1|_{\Sigma}\otimes V_2^{-1}|_{\Sigma})\\
n_{\tilde{X}}&= h^0(\Sigma,K_{\Sigma}^{1/2}\otimes V_1^{-1}|_{\Sigma}\otimes V_2|_{\Sigma})\,,
\end{split}\end{equation}
where $K_{\Sigma}^{1/2}$ is the spin bundle on $\Sigma$.

As discussed in \cite{Heckman:2008es}, nontrivial superpotential terms can be generated by D3-instantons which wrap the 4-cycle $S$. Such an instanton houses its own gauge group $U(1)_{\text{inst}}$ and we must further sum over all possible choices of supersymmetric bundle $V_{\text{inst}}$ 
and hence all possible supersymmetric instanton configurations. As pointed out and discussed in much detail in v2 of \cite{Heckman:2008es} it is important that in order for the D3-instanton to contribute, the flux on the instanton has to be trivializable. I.e. it is a two-form, which on the 4-cycle $S$ is dual to a 2-cycle, which however, is trivial in homology of the full geometry. Thus, the sharper statement is, that we need to sum over all such trivializable flux configurations on the D3-instanton, for a fixed 7-brane world-volume flux. We will assume this trivializability, which is a global condition, throughout the paper.

 Rather than parametrize these configurations by $V_{\text{inst}}$, however, we will find that it is more convenient to use the bundle ${\cal{L}}$ defined by
\begin{equation}{\cal{L}}\equiv V_{\text{inst}}\otimes V_1^{-1}\,.
\end{equation}
For a given choice of ${\cal{L}}$ the superpotential interactions that can be generated, if any, are determined by the number of fermi zero modes which connect it to the two D7-branes.  This counting is reviewed in Appendix \ref{app:instantons}, where we also observe that the 3-7 and 7-3 zero mode structure implies that for $\Sigma$ of genus $g_{\Sigma}\le 1$ only three types of superpotential coupling can be generated.  These take the form
\begin{equation}W^{(1)}\sim X^p,\qquad W^{(2)}\sim \tilde{X}^q,\quad \text{and}\quad W^{(3)}\sim X\tilde{X}\label{possWs}\end{equation}
with $0\le p,q\le N-2$ for $S=dP_N.$
 For higher genus $g_{\Sigma}>1$ additional couplings may be possible but we restrict to $g_{\Sigma}\le 1$ in this note for simplicity.

In order to use this system for model building, it is useful to determine when these various couplings can be generated and, if so, what bundle choices do the job.  The collection of bundles that we have to choose from, however, is restricted to those which are supersymmetric and this in turn depends on our choice for the K\"ahler form $J_S$ on $S$.  More specifically, given $J_S$ we must sum over instanton bundles $V_{\text{inst}}$ which satisfy
\begin{equation}\int_S\,J_S\wedge c_1(V_{\text{inst}})=0 \,.
\end{equation}
In this note, it will be useful for us to assume that $J_S$ takes the so-called large volume form{\footnote{We use this condition for exactly one purpose, namely ruling out the possibility that bundles ${\cal{L}}$ with $c_1({\cal{L}})\cdot H\ne 0$ yield nontrivial superpotential contributions. We will further make use of the standard notation for exceptional divisors for $dP_n$ surfaces: $E_i \cdot E_j =- \delta_{ij}$, $H^2 =1$, and $H\cdot E_i =0$.}}
\begin{equation}J_S = AH - \sum_{i=1}^N B_i E_i\,,\qquad\qquad A,B_i>0\,,\qquad A\gg B_i \qquad \forall i  \,.
\label{KLV}\end{equation}


\subsection{Generating $W^{(1)}\sim X^p$}

Let us turn now to the three types of superpotential couplings in \eqref{possWs}.  In Appendix \ref{app:instantons}, we demonstrate that a necessary condition for generation of superpotential terms of the form $W^{(1)}\sim X^p$ is that ${\cal{L}}$ takes the form
\begin{equation}
{\cal{L}}^{(1)}_p={\cal{O}}\left(E_1-\sum_{j=2}^{p+2}E_j\right) \,,
\label{L1p}\end{equation}
up to permutations of the exceptional divisors $E_i$.  This is a condition which follows from counting fermi zero modes between the instanton and D7$_1$ and hence is completely independent of our choice of class for $\Sigma$.  For ${\cal{L}}$ of the form \eqref{L1p} a sufficient condition for the generation of a superpotential coupling $W^{(1)}\sim X^p$ is then
\begin{equation}
\text{deg}\left({\cal{L}}^{(1)}_p|_{\Sigma}\right)=-\text{deg}\left(V_1|_{\Sigma}\otimes V_2^{-1}|_{\Sigma}\right)-p \,.
\label{L1pgencond}\end{equation}

\subsection{Generation $W^{(2)}\sim \tilde{X}^q$}

A necessary condition for generating $W^{(2)}\sim \tilde{X}^q$, on the other hand, is that ${\cal{L}}$ takes the form
\begin{equation}
{\cal{L}}^{(2)}_m={\cal{O}}\left(-E_1+\sum_{j=2}^{m+2}E_j\right) \,,
\label{L2m}\end{equation}
up to permutations of the exceptional divisors $E_i$.  As with \eqref{L1p}, this is a condition which follows from counting fermi zero modes between the instanton and D7$_1$ and hence does not depend on the class of $\Sigma$.  For ${\cal{L}}$ of the form \eqref{L2m} a sufficient condition which guarantees the generation of $W^{(2)}\sim \tilde{X}^m$ is then given by
\begin{equation}\text{deg}\left({\cal{L}}^{(2)}_m|_{\Sigma}\right)=-\text{deg}\left(V_1|_{\Sigma}\otimes V_2^{-1}|_{\Sigma}\right)+m \,.
\end{equation}

\subsection{Generating $W^{(3)}\sim X\tilde{X}$}

Finally, we turn our attention to the superpotential coupling $W^{(3)}\sim X\tilde{X}$.  The 3-7 and 7-3 zero mode structure implies that such a coupling cannot arise for $\Sigma$ of genus 0 so we restrict our attention to the case $g_{\Sigma}=1$.
A necessary condition for generating $W^{(3)}$ is that ${\cal{L}}$  takes the form
\begin{equation}{\cal{L}}^{(3)}={\cal{O}}\left(E_i+E_j-E_k-E_{\ell}\right)\qquad i,j,k,\ell\text{ all distinct}\,.
\label{L3}\end{equation}
A further condition which guarantees generation of $W^{(3)}$ is then simply
\begin{equation}
\text{deg}\left({\cal{L}}^{(3)}|_{\Sigma}\right)=-\text{deg}\left(V_1|_{\Sigma}\otimes V_2^{-1}|_{\Sigma}\right) \,.
\label{L3cond}\end{equation}

This exhausts all choices for ${\cal{L}}$ which can generate nontrivial superpotential couplings when $g_{\Sigma}\le 1$.

\subsection{Coupling to $SU(5)$ F-theory GUTs and Trouble with $\mu$ Term}

In the rest of this note, we will turn to a series of applications of these results in order to demonstrate that this simple configuration of two intersecting D7-branes comprises a useful toolbox for model building with intersecting 7-branes.  Before getting ahead of ourselves, though, we first note that, when this system is incorporated into an $SU(5)$ F-theory GUT \cite{Beasley:2008dc,Beasley:2008kw}, it necessitates the introduction of a second matter curve, $\Sigma_{GUT}$, in figure \ref{fig:Pol} along which the D7 wrapping $S$ intersects the GUT branes.  This extra matter curve is not only the source of some of the GUT-charged matter which participates in the coupling \eqref{oursup}, it can also give rise to extra fermi zero modes between the GUT branes and D3-instanton which can alter or even possibly eliminate our instanton-generated superpotential terms altogether.

When $\Sigma_{GUT}=\mathbb{P}^1$, one simple way to ensure the absence of such zero modes for a given instanton configuration is to choose $\Sigma_{GUT}$ so that both the hypercharge line bundle on $S_{GUT}$ and the $U(1)_{\text{inst}}$ bundle on the instanton restrict trivially there.  As such, if we want to obtain a specific superpotential coupling, we need only make sure that this condition holds for at least one of the choices for a trivializable $U(1)_{\text{inst}}$ bundle that can generate it.

There is one important situation in which we cannot get rid of these extra zero modes, however, and that is when there are Higgs fields localized on $\Sigma_{GUT}$.  In order to achieve doublet-triplet splitting in $SU(5)$ F-theory GUTs, one turns on a nontrivial hypercharge bundle which restricts trivially to the matter curves which house quarks and leptons but nontrivially to matter curves which house the Higgs fields \cite{Beasley:2008kw}.  When counting fermi zero modes between the GUT branes and instantons, doublets and triplets carry different hypercharge, though, so it is impossible to remove all of them regardless of how we choose the class of $\Sigma_{GUT}$ or the bundle $U(1)_{\text{inst}}$.  For this reason, nonzero couplings which arise from D3-instantons wrapping Higgs matter branes in $SU(5)$ F-theory GUTs will always include some $SU(2)\times SU(3)$ invariant combination of the corresponding Higgs doublet and possibly also doublet or triplet modes from its Kaluza-Klein tower{\footnote{In fact, most, if not all, of these couplings break hypercharge so we are essentially forced to assume that the K\"ahler form is constrained in such a way that the instantons which generate them are nonsupersymmetric.}}.  

Because the $\mu$ term arises from the intersection of two Higgs matter curves in this setup, we therefore conclude that it is not possible for D3-instantons to provide the bosonic expectation value that one needs to generate it.  All of the other scales that we seek to generate, however, arise from triple intersections with at least one matter curve to which the $U(1)_Y$ bundle in $SU(5)$ F-theory GUTs restricts trivially.  That the $\mu$ term is singled out is actually quite interesting, as it suggests that perhaps it should not be an independent scale but rather should be generated in connection with one of the others.
One example of this, for instance, is the gauge mediation framework of \cite{Marsano:2008jq} where the generation of $\mu$ is connected with supersymmetry-breaking and hence of the same order as the supersymmetry-breaking scale.


\section{Linear Superpotential: Supersymmetry Breaking and the $\rho$ Coupling}
\label{sec:linear}

We begin by considering linear superpotential terms and their
application to supersymmetry breaking
 and the $\rho$ coupling \eqref{rhoterms}.
 Throughout this section, we shall assume for simplicity that $\Sigma=\mathbb{P}^1$.  In this case, the massless chiral multiplets localized on $\Sigma$ are comprised entirely of $X$'s or of $\tilde{X}$'s.  Without loss of generality, then, we suppose that the degree of $V_1|_{\Sigma}\otimes V_2^{-1}|_{\Sigma}$ is nonnegative
so that
\begin{equation}\begin{split}
n_{X} &= \text{deg}\left(V_1|_{\Sigma}\otimes V_2^{-1}|_{\Sigma}\right)\ge 0\\
n_{\tilde{X}}&=0 \,.
\end{split}\label{nxnxt}\end{equation}
Superpotential terms of degree $p$ can then be generated by instanton configurations with ${\cal{L}}$ of the form ${\cal{L}}^{(1)}_p$ in \eqref{L1p},
which further satisfy the condition \eqref{L1pgencond}, which we rewrite as
\begin{equation}
\deg\left({\cal{L}}|_{\Sigma}\right)=-n_X-p \,.
\label{pcond}\end{equation}


\subsection{Supersymmetry Breaking}

\subsubsection{Setup}

It has already been established in \cite{Heckman:2008es} that one can engineer a single chiral field $X$ and generate for it a SUSY-breaking Polonyi superpotential using D3-instantons.  To review this construction,
we consider the case $n_X=1$ and seek to generate a superpotential that is purely linear in $X$.  For $S=dP_N$ with $N\ge 3$, one choice for $\Sigma$, which allows some of the bundles ${\cal{L}}^{(1)}_p$ with $p=1$ to contribute is simply
\begin{equation}\Sigma = H-E_1-E_2\,.\end{equation}
In this case, the degree of ${\cal{L}}^{(1)}_p|_{\Sigma}$ is bounded from below
\begin{equation}
\deg\left({\cal{L}}^{(1)}_p\right)\ge -2\,,
\end{equation}
so that \eqref{pcond} admits no solutions for $p>1$.  For $p=1$, however, there are configurations ${\cal{L}}^{(1)}_p$ which contribute, namely those of the form
\begin{equation}
{\cal{L}}={\cal{O}}\left(E_i-E_1-E_2\right)\qquad i>3\,,
\end{equation}
so that linear superpotential couplings are indeed generated.

\subsubsection{O'Raifeartaigh Model}

In addition to the linear superpotential $W\sim X$ for the massless field $X$, the superpotential of this model at the KK scale also includes couplings to the various KK modes in the problem.  In particular, there are KK modes for both $X$ and $\tilde{X}$, and the $U(1)$ "adjoint" field $\Phi$, which is a (0,1) form on $S$.  Including also the instanton-generated linear terms, the full superpotential including these tree-level couplings takes the form{\footnote{We neglect the instanton-generated term $\tilde{\mu} X_{KK}$ involving the KK mode $X_{KK}$ because it plays no essential role here.}}
\begin{equation}
W=  F_X X + \lambda X {\tilde X}_{KK}\Phi_{KK}+m_1X_{KK}{\tilde X}_{KK}+m_2\Phi_{KK}^2  \,,
\end{equation}
where both masses $m_1$ and $m_2$ are of the order $\mathcal{O}(M_{KK})$.
At low scales, one typically drops terms involving the KK modes
since they may be integrated out in a supersymmetric fashion, i.e. by solving the corresponding  F-term equations.
Note, however, that with KK modes included the superpotential that we obtain is essentially
the O'Raifeartaigh model discussed in \cite{Shih:2007av}.  That we obtain a Polonyi superpotential, when KK modes are thrown away, is simply the well-known fact that the O'Raifeartaigh superpotential reduces to a Polonyi one for the pseudo-modulus at low energies.  However, the analysis of \cite{Shih:2007av} establishes that including 1-loop corrections from integrating out the KK modes leads to a Coleman-Weinberg potential that lifts the pseudo-moduli space
as long as
\begin{equation}
\frac{\lambda F_X}{2m_1m_2}\sim \frac{\lambda F_X}{M_{KK}^2}<1 \,.
\end{equation}
This is of course the regime of interest for us.   The actual value of $\langle X\rangle$ at the minimum depends on the ratio  $r=2 m_2/m_1$.  This expectation value is vanishing when $r<2.11$ but becomes nonzero for larger values of $r$.  We would prefer the latter case because $\langle X\rangle$ is responsible for giving mass to the gauge messengers when this model is used to implement gauge mediation.  Note that in our local construction, we only know about the scale of the masses, which are $m_i\sim M_{KK}$.  To ensure that $\langle X\rangle$ is indeed nonzero, we would need to know the numerical coefficients more precisely.

Note that, as pointed out in \cite{Marsano:2008jq}, a different contribution, which arises from integrating out the massive $U(1)$ gauge boson, generates a quartic correction to the K\"ahler potential for $X$ that also generically lifts the pseudo-moduli space \cite{ArkaniHamed:1998nu}.  This effect will compete with the one described here from integrating out KK modes.  Which one dominates is difficult to determine, however, because the mass of the $U(1)$ gauge boson is determined by its couplings to closed string modes and hence, as far as we know, can take a wide range of values.


\subsection{$\rho$ Coupling}
\label{subsec:linearmu}

Alternatively, we can use instanton-induced linear superpotential couplings to generate expectation values, which, in turn, can be used to introduce $\rho$ couplings \eqref{rhoterms} which implement the gauge mediation scenario of \cite{Marsano:2008jq} in which the Higgs and messenger sectors are coupled together at $SU(7)$ enhancement points.  The basic idea is to start with bundles $V_1$ and $V_2$ such that $n_X=n_{\tilde{X}}=0$.  In that case, one only has Kaluza-Klein modes $X_I$ and $\tilde{X}_I$ localized on $\Sigma$ with superpotential
\begin{equation}
W\sim \sum_I c_I M_{KK} X_I \tilde{X}_I\,,
\end{equation}
for some suitable coefficients $c_I$.  If we can succeed in using instantons to generate terms that are linear in $X$ and $\tilde{X}$ then the resulting superpotential will take the form
\begin{equation}
W\sim \sum_I \left(m^2\left[d_I X_I + e_I \tilde{X}_I\right]+ c_I M_{KK}X_I\tilde{X}_I\right)\,,
\end{equation}
where $m^2$ is the instanton-generated scale and $d_I,e_I$ are numerical coefficients, which we can very roughly assume to be ${\cal{O}}(1)$ or so.  The F-term equations for this system yield nonzero expectation values
\begin{equation}
X_I = -\frac{e_I}{c_I}\left(\frac{m^2}{M_{KK}}\right)\,,\qquad
\tilde{X}_I = -\frac{d_I}{c_I}\left(\frac{m^2}{M_{KK}}\right)\,.\end{equation}
These can then be used to generate a direct $\rho$ coupling \eqref{rhoterms} between the Higgs and messenger sectors in models of gauge mediation \cite{Marsano:2008jq} through the interaction
\begin{equation}W\sim \lambda_I X_I H\bar{f}\,.
\end{equation}
and the analogous one involving $\bar{H}$ and $f$.  Neglecting contributions of the various numerical coefficients, we see that the rough scale of $\rho$ is set by the ratio $m^2/M_{KK}$ and hence is exponentially suppressed in this case.

Returning now to the condition \eqref{pcond}, we see that in the case of interest, $n_X=0$, bundles ${\cal{L}}^{(1)}_p$ in \eqref{L1p} with $p=1$ can contribute, giving rise to $W_{inst}\sim X_I,$ if we make the particularly simple choice
\begin{equation}\Sigma = E_1\,.\end{equation}
Furthermore, with this choice the restrictions of the ${\cal{L}}^{(1)}_p$ in \eqref{L1p} to $\Sigma$ have   degrees which are bounded from below
\begin{equation}
\deg\left({\mathcal{L}}^{(1)}_p|_{\Sigma}\right)\ge -1\,,\end{equation}
and hence cannot satisfy \eqref{pcond} for $p>1$.  This means that no
$p>1$ powers of $X_I$ can be generated.
Similarly, ${\mathcal L}^{(2)}_m$ with $m=1$ contributes for this choice of $\Sigma,$
giving rise to $W_{\text{inst}}\sim {\tilde X}_I.$
Meanwhile, no $m>1$ powers of ${\tilde X}_I$ can be generated.


\section{Quadratic Couplings: Neutrino Masses and $\rho$ Coupling}
\label{sec:quadratic}

We now turn to the generation of quadratic superpotential couplings.  As we shall see, such models can be applied to engineer a Majorana mass for the right-handed neutrino or, alternatively, to generate the nonzero expectation value needed to obtain a $\mu$ term in the Higgs sector.  As in the previous section, we make the simplifying assumption here that $\Sigma=\mathbb{P}^1$ and consider, without loss of generality, the situation where the degree of $V_1|_{\Sigma}\otimes V_2^{-1}|_{\Sigma}$ is nonnegative.  In particular, this means that $n_X\ge 0$ and $n_{\tilde{X}}=0$ as in \eqref{nxnxt}.

\subsection{Right-handed Neutrino Mass}

To study applications to the neutrino sector, we now suppose that $n_X=1$ and identify the chiral multiplet $X$ with the multiplet which contains the right-handed neutrino, $N_R$.  The desired quadratic superpotential couplings can be generated by instanton configurations of the form ${\cal{L}}^{(1)}_p$ in \eqref{L1p} with $p=2$
\begin{equation}{\cal{L}}^{(1)}_2={\cal{O}}\left(E_i-E_j-E_k-E_{\ell}\right)\qquad i,j,k,\ell\text{ all distinct}\,,\end{equation}
which further satisfy \eqref{pcond}, i.e.
\begin{equation}\deg\left({\cal{L}}|_{\Sigma}\right)=-n_X-2\,.
\end{equation}
One possible solution arises if we take $S=dP_N$ with $N\ge 4$ and further choose $\Sigma$ as
\begin{equation}
\Sigma = 2H - E_1 - E_2 - E_3\,,
\end{equation}
since in this case we get a nonzero superpotential contribution from ${\cal{L}}$ of the form
\begin{equation}
{\cal{L}}_i= {\cal{O}}\left(E_i-E_1-E_2-E_3\right)\qquad i>3\,.
\label{NeutLi}\end{equation}
For this choice of $\Sigma$, it is easy to see that none of the ${\cal{L}}^{(1)}_p$ of \eqref{L1p} with $p>2$ satisfy \eqref{pcond} so no superpotential terms of degree greater than 2 can be generated.  On the other hand, there are bundles ${\cal{L}}^{(1)}_p$ with $p=1$ which satisfy \eqref{L1pgencond}, namely
\begin{equation}{\cal{L}}={\cal{O}}\left(E_j-E_1-E_2\right),\quad {\cal{O}}\left(E_j-E_2-E_3\right),\quad {\cal{O}}\left(E_j-E_1-E_2\right)\quad j>3\,.
\label{Neutother}\end{equation}
For $S=dP_4$, it is easy to see that none of these can be simultaneously supersymmetric with the bundle ${\cal{L}}_i$ in \eqref{NeutLi} which generates the quadratic coupling.  This follows from the fact that the exceptional divisors $E_j$ have nonzero area.  On the other hand, for $S=dP_N$ with $N>4$ it is possible that one of the bundles in \eqref{Neutother} is simultaneously supersymmetric with a bundle from \eqref{NeutLi} but this requires a further tuning of the K\"ahler form $J_S$.  In order to generate only a mass term for $X$, we must assume that this further tuning does not take place.

\subsection{$\rho$ Coupling}

Note, however, that if $J_S$ is such that there in fact do exist bundles of the form \eqref{NeutLi} and \eqref{Neutother} which are simultaneously supersymmetric, we obtain instead a superpotential of the form
\begin{equation}W\sim m^2 X + \tilde{m} X^2\,,\end{equation}
where $m^2$ and $\tilde{m}$ each contain an instanton-generated exponential suppression factor.  This would be disastrous in a model where $X$ is identified with the right-handed neutrino because $X$ picks up a nonzero expectation value.  On the other hand, we have already seen that a model with nonzero $\langle X\rangle$ can be useful for other purposes, namely the generation of the $\rho$ coupling \eqref{rhoterms} which appears in one of the gauge mediation scenarios of \cite{Marsano:2008jq}.  Note that the instanton suppression factor essentially cancels out from the ratio $m^2/\tilde{m}$ so that the expectation value for $X$ obtained in this manner sits naturally near the Kaluza-Klein scale
\begin{equation}\langle X\rangle \sim M_{KK}\,.\end{equation}


\section{Mixed Couplings and Fayet}
\label{sec:mixed}

We now turn our attention to the generation of mixed couplings of the form $X\tilde{X}$.  Such a coupling is in fact precisely what one needs to engineer a Fayet model of supersymmetry breaking.  As discussed in section \ref{sec:setup}, one cannot simultaneously have massless $X$ and $\tilde{X}$ when $\Sigma=\mathbb{P}^1$ so we consider instead the case of $\Sigma=T^2$.  To obtain $n_X=n_{\tilde{X}}=1$, we further require that
\begin{equation}\deg\left(V_1|_{\Sigma}\otimes V_2^{-1}|_{\Sigma}\right)=0\,.\end{equation}

\subsection{Superpotential Couplings}

The desired superpotential coupling $X\tilde{X}$ can be generated by bundles ${\cal{L}}^{(3)}$ of the form \eqref{L3} which further satisfy \eqref{L3cond}, a condition that we can rewrite here as
\begin{equation}\deg\left({\cal{L}}|_{\Sigma}\right)=0\,.
\end{equation}
If we take $S=dP_N$ with $N\ge 4$ then one choice for $\Sigma$ which leads to generation of the coupling $X\tilde{X}$ is given by
\begin{equation}\Sigma = 3H - E_1 - E_2\,.\label{SigFayet}\end{equation}
since then we get nonzero contributions from bundles of the form
\begin{equation}{\cal{L}}^{(X\tilde{X})}_{ij}={\cal{O}}\left(E_1- E_2+E_i-E_j\right)\quad i\ne j,\quad i,j>2\,,
\label{FXXt1}\end{equation}
and
\begin{equation}{\cal{L}}^{(X\tilde{X})}_{ij}={\cal{O}}\left(-E_1+ E_2+E_i-E_j\right)\quad i\ne j,\quad i,j>2\,.
\label{FXXt2}\end{equation}

In addition to $X\tilde{X}$, it is possible that further superpotential couplings are generated.  Bundles which generate $X^p$ arise from ${\cal{L}}$ of the form ${\cal{L}}^{(1)}_p$ in \eqref{L1p}, which also satisfy \eqref{L1pgencond}
\begin{equation}\deg\left({\cal{L}}|_{\Sigma}\right)=-p\,.\end{equation}
For our choice \eqref{SigFayet} of $\Sigma$, however, the degree of ${\cal{L}}^{(1)}_p|_{\Sigma}$ is bounded from below
\begin{equation}
\deg\left({\cal{L}}|_{\Sigma}\right)\ge -2\,,
\end{equation}
so no terms $X^p$ of degree $p>2$ can be generated.  Terms with $p=1$ can be generated, though, by ${\cal{L}}$ of the form
\begin{equation}{\cal{L}}^{(X)}_{ija}={\cal{O}}\left(E_i-E_j-E_a\right)\qquad a=1,2\qquad i\ne j,\quad i,j>2\label{FX} \,,
\end{equation}
while terms with $p=2$ can be generated by
\begin{equation}{\cal{L}}^{(X^2)}_{ij}={\cal{O}}\left(E_i-E_1-E_2-E_j\right)\qquad i\ne j\quad i,j>2\,.\label{FXX}\end{equation}

The story for superpotential couplings involving $\tilde{X}$ is similar.  Couplings of the form $\tilde{X}^p$ with $p>2$ are not generated while a linear coupling, $\tilde{X}$, can be obtained from configurations with
\begin{equation}{\cal{L}}^{(\tilde{X})}_{aij}={\cal{O}}\left(-E_i+E_j+E_a\right)\quad a=1,2\quad i\ne j\quad i,j>2\label{FXt}\end{equation}
and a quadratic coupling $\tilde{X}^2$ can arise from
\begin{equation}
{\cal{L}}^{(\tilde{X}^2)}_{ij}=
{\cal{O}}\left(-E_i+E_1+E_2+E_j\right)\, \qquad i\ne j \qquad i,j >2\,.\label{FXtXt}\end{equation}

For our choice \eqref{SigFayet} of $\Sigma$, then, we see that the most general superpotential, which can be generated by D3-instantons takes the form
\begin{equation}W\sim aX + b\tilde{X} + \frac{c}{2}X^2 + \frac{d}{2}\tilde{X}^2 + mX\tilde{X} \,.\end{equation}
Whether the couplings $a,b,c,d,m$ are actually generated depends of course on which if any of the bundles \eqref{FX}, \eqref{FXt}, \eqref{FXX}, \eqref{FXtXt}, \eqref{FXXt1}, and \eqref{FXXt2} are supersymmetric.  To realize a mixed Fayet-type superpotential coupling $m\ne 0$, it is always necessary to assume that $J_S$ is such that at least one of \eqref{FXXt1} or \eqref{FXXt2} is supersymmetric.  Within the space of $J_S$, which satisfy this condition, though, the bundles needed to generate nonzero $a,b,c,d$ will in general not be supersymmetric.  Further relations must be imposed in order to generate some of these terms simultaneously with $m$.  For simplicity, we shall assume the more generic situation where $a=b=c=d=0$.

\subsection{Nonzero FI-parameter and Supersymmetry Breaking}

The Fayet model of supersymmetry breaking relies on more than the mixed superpotential coupling $X\tilde{X}$.  In addition, we need to have a nonzero Fayet-Iliopolous parameter for one of the $U(1)$ gauge fields in the problem.  This is actually easy to achieve, for instance, by choosing the gauge bundle $V_2$ on D7$_2$ to be nonsupersymmetric
\begin{equation}\int_{S_2}J_2\wedge c_1(V_2)=\xi^{(2)}_{FI}\ne 0 \,,
\end{equation}
where $S_2$ is the 4-cycle wrapped by D7$_2$ and  $J_2$ is its K\"ahler form.  The object $\xi^{(2)}_{FI}$ is nothing more than a Fayet-Iliopolous parameter for $U(1)_2$.  As such, we have not only the Fayet superpotential
\begin{equation}W_{Fayet}\sim mX\tilde{X}\,,\end{equation}
but also the  $D$-terms with
\begin{equation}D_1\sim |X|^2-|\tilde{X}|^2\,,\qquad D_2\sim |\tilde{X}|^2-|X|^2-\xi^{(2)}_{FI}\,.\end{equation}
As discussed for instance in \cite{Aharony:2007db}, this model breaks supersymmetry at an exponentially small scale $m\sqrt{\xi^{(2)}_{FI}}$ determined by the instanton-generated coupling $m$.


\section*{Acknowledgements}

We are grateful to J.~Heckman and C.~Vafa for useful discussions and collaboration on the earlier work \cite{Heckman:2008es}.  The work of JM and SSN was supported by Caltech John A. McCone Postdoctoral Fellowships.  The work of NS was supported in part by the DOE-grant DE-FG03-92-ER40701.

\appendix

\section{General Conditions for Generating Superpotentials with D3 Instantons}
\label{app:instantons}

In this Appendix, we study the generation of superpotential couplings by D3-instantons in the simple setup of a pair of intersecting D7-branes.  More specifically, one D7-brane, denoted D7$_1$, wraps a del Pezzo surface $S=dP_N$ while the second, D7$_2$, intersects $S$ along a curve $\Sigma$.  We consider here the effects of BPS D3-instantons, which also wrap $S$.  This setup is depicted in figure \ref{fig:Pol}.  We denote the gauge groups on the D7$_i$ by $U(1)_i$ and the gauge group on the D3-instanton by $U(1)_{\text{inst}}$.  Supersymmetric gauge bundles $V_i$ and $V_{\text{inst}}$ associated to these gauge groups on the worldvolume satisfy the condition
\begin{equation}\int \, J \wedge c_1(V)=0\,,
\end{equation}
where the integral is performed over the compact part of the worldvolume and $J$ is the corresponding K\"ahler form.  In all that follows, we will assume that the K\"ahler form on $S$, denoted $J_S$, takes the large volume form \eqref{KLV}
\begin{equation}J_S = AH - \sum_{i=1}^N B_i E_i\,,\qquad A,B_i>0\,,\qquad A\gg B_i\,.
\label{ourkahler}\end{equation}

To study instanton-generated superpotentials we need to count fermi zero modes connecting the instanton to the various D7-branes.  Using $n_{pqr}$ to denote the number of fermi zero modes with charges $(p,q,r)$ under $U(1)_1\times U(1)_2\times U(1)_{\text{inst}}$ we recall the general formulae of \cite{Heckman:2008es}
\begin{equation}\begin{split}
n_{+-0}&=h^0(\Sigma,K_{\Sigma}^{1/2}\otimes V_1|_{\Sigma}\otimes V_2^{-1}|_{\Sigma})\\
n_{-+0}&=h^0(\Sigma,K_{\Sigma}^{1/2}\otimes V_1^{-1}|_{\Sigma}\otimes V_2|_{\Sigma})\\
n_{+0-}&=-\chi(S,V_1\otimes V_{\text{inst}}^{-1})\\
n_{-0+}&=-\chi(S,V_1^{-1}\otimes V_{\text{inst}})\\
n_{0+-}&=h^0(\Sigma,K_{\Sigma}^{1/2}\otimes V_2|_{\Sigma}\otimes V_{\text{inst}}^{-1}|_{\Sigma})\\
n_{0-+}&= h^0(\Sigma,K_{\Sigma}^{1/2}\otimes V_2^{-1}|_{\Sigma}\otimes V_{\text{inst}}|_{\Sigma})\,.
\end{split}\end{equation}
In what follows, it will be important that $K_{\Sigma}={\cal{O}}(-2)$ for $\Sigma=\mathbb{P}^1$ and $K_{\Sigma}={\cal{O}}$ for $\Sigma=T^2$.  We shall restrict for simplicity to only these two cases.

A fairly obvious pair of conditions that a given instanton configuration must satisfy in order to yield a nontrival contribution to the superpotential consists of
\begin{equation}n_{+0-}=n_{0-+}\qquad\text{and}\qquad n_{-0+}=n_{0+-}\,.\end{equation}
In almost all situations with $\Sigma$ of genus 0 or 1, only one of $n_{0-+}$ and $n_{0+-}$ is nonzero.  The only exception to this corresponds to the special situation where $\Sigma=T^2$ and $V_2|_{\Sigma}\otimes V_{\text{inst}}^{-1}|_{\Sigma}={\cal{O}}$, in which case $n_{0+-}=n_{0-+}=1$.  We thus have three possibilities to consider, namely $n_{0-+}=0$, $n_{0+-}=0$, and this special case $n_{0+-}=n_{0-+}=1$.  For each situation, we will classify all supersymmetric instanton bundles that can yield nontrivial superpotential couplings and the form of the couplings that they can generate.  
Furthermore, as discussed in section \ref{sec:setup}, the flux associated to $V_{\rm inst}$ on the D3-instanton has to be trivializable. 
In what follows, it will be useful to parametrize instanton configurations not by $V_{\text{inst}}$ but rather by the bundle
\begin{equation}
{\cal{L}}=V_{\text{inst}}\otimes V_1^{-1}\,.
\end{equation}
When solving for ${\cal{L}}$ we will often use the ansatz
\begin{equation}
{\cal{L}}={\cal{O}}\left(b_0H + \sum_{i=1}^N b_i E_i\right)\,.
\label{Lansatz}\end{equation}
Given the assumed from of $J_S$ in \eqref{ourkahler}, supersymmetric ${\cal{L}}$ are built from $b_0$ and $b_i$ satisfying
\begin{equation}Ab_0+\sum_{i=1}^N B_i b_i=0\,.\end{equation}
This, together with the assumption that $A\gg B_i$, means that if $b_0\ne 0$, there must exist at least one $b_i$, say $b_{i_0}$, which satisfies
\begin{equation}|b_{i_0}| \gg |b_0|\,.\label{b0cond}\end{equation}
We now turn to a study of the three cases enumerated above.


\subsection{Case 1: $n_{0-+}=0$ and $n_{0+-}=p>0$}

This case arises when $V_2$ and $V_{\text{inst}}$ satisfy
\begin{equation}\text{deg}\left(\left[V_2\otimes V_{\text{inst}}^{-1}\right]|_{\Sigma}\right)=p\,.
\end{equation}
We can rewrite the above condition in terms of ${\cal{L}}$ as
\begin{equation}\text{deg}\left({\cal{L}}|_{\Sigma}\right)=-\text{deg}\left(\left[V_1\otimes V_2^{-1}\right]|_{\Sigma}\right)-p\,.\end{equation}
Note that the object $\text{deg}([V_1\otimes V_2^{-1}]|_{\Sigma})$ is precisely what determines the spectrum of chiral massless fields localized on $\Sigma$.

To obtain a nontrivial superpotential contribution, we must also have
\begin{equation}n_{-0+}=p\qquad\text{and}\qquad n_{+0-}=0\,.\end{equation}
Expressed in terms of the parameters in our ansatz \eqref{Lansatz}, these equations take the form
\begin{equation}3b_0+\sum_{i=1}^Nb_i = -p\qquad\text{and}\qquad b_0^2-\sum_{i=1}^N b_i^2 = -p-2\,.
\label{c1cond}\end{equation}
Eliminating $p$ from \eqref{c1cond} then yields the condition
\begin{equation}\left(b_0-\frac{3}{2}\right)^2 + \frac{N-1}{4}=\sum_{i=1}^N \left(b_i+\frac{1}{2}\right)^2\,.
\label{c1bcond}\end{equation}
No solutions to \eqref{c1bcond} with $b_0\ne 0$ satisfy \eqref{b0cond} so we conclude that the only supersymmetric bundles ${\cal{L}}$ which satisfy \eqref{c1cond} have $b_0=0$.  In that case, \eqref{c1bcond} becomes the fairy simple equation
\begin{equation}
\sum_{i=1}^N\left(b_i+\frac{1}{2}\right)^2=2+\frac{N}{4}\,.
\end{equation}
Solutions to this equation are easy to enumerate.  One coefficient, say $b_{i_0}$, must satisfy $b_{i_0}=1$ and the rest must be either 0 or -1{\footnote{There is one additional class of solutions in which $b_{i_0}=-2$ and the rest are either 0 or -1.  Such solutions cannot be supersymmetric since the volumes of all exceptional divisors are positive.}}.  Returning to \eqref{c1cond}, we then see that the number of $b_i$'s which take the value $-1$ must be $p+1$.  As such, we find that for a given value of $p$ there is exactly one supersymmetric choice of ${\cal{L}}$ up to permutation of the $E_i$ which satisfies \eqref{c1cond}, namely
\begin{equation}{\cal{L}}^{(1)}_p={\cal{O}}\left(E_1-\sum_{i=2}^{p+2}E_i\right)\,,
\end{equation}
provided $p\le N-2$.

\subsection{Case 2: $n_{0+-}=0$ and $n_{0-+}=m>0$}

This case arises when $V_2$ and $V_{\text{inst}}$ satisfy
\begin{equation}\text{deg}\left(\left[V_{\text{inst}}\otimes V_2^{-1}\right]|_{\Sigma}\right)=m \,.
\end{equation}
We can rewrite the above condition in terms of ${\cal{L}}$ as
\begin{equation}\text{deg}\left({\cal{L}}|_{\Sigma}\right)=-\text{deg}\left(\left[V_1\otimes V_2^{-1}\right]|_{\Sigma}\right)+m
\,.
\end{equation}
Note that the object $\text{deg}([V_1\otimes V_2^{-1}]|_{\Sigma})$ is what determines the spectrum of chiral massless fields localized on $\Sigma$.

To obtain a nontrivial superpotential contribution, we must also have
\begin{equation}n_{-0+}=0\qquad\text{and}\qquad n_{+0-}=m\,.\end{equation}
Expressed in terms of the parameters in our ansatz \eqref{Lansatz}, these equations take the form
\begin{equation}3b_0+\sum_{i=1}^N b_i = m\qquad\text{and}\qquad b_0^2-\sum_{i=1}^N b_i^2 = -m-2 \,.
\label{c2cond} \end{equation}
Eliminating $m$ from \eqref{c2cond} then yields the condition
\begin{equation}\left(b_0+\frac{3}{2}\right)^2 +\frac{N-1}{4}=\sum_{i=1}^N\left(b_i-\frac{1}{2}\right)^2 \,.
\label{c2bcond}\end{equation}
No solutions to \eqref{c2bcond} with $b_0\ne 0$ satisfy \eqref{b0cond} so we conclude that the only supersymmetric bundles ${\cal{L}}$ which satisfy \eqref{c2cond} have $b_0=0$.  In that case, \eqref{c2bcond} becomes the fairly simple equation
\begin{equation}\sum_{i=1}^N \left(b_i-\frac{1}{2}\right)^2=2+\frac{N}{4}\,.
\end{equation}
Solutions to this equation are easy to enumerate.  One coefficient, say $b_{i_0}$, must satisfy $b_{i_0}=-1$ and the rest must be either 0 or 1{\footnote{There is one additional class of solutions in which $b_{i_0}=2$ and the rest are either 0 or 1.  Such solutions cannot be supersymmetric since the volumes of exceptional divisors are positive.}}.  Returning to \eqref{c2cond}, we then see that the number of $b_i$'s which take the value 1 must be $m+1$.  As such, we find that for a given value of $m$ there is exactly one supersymmetric choice of ${\cal{L}}$ up to permutation of the $E_i$ which satisfies \eqref{c2cond}, namely
\begin{equation}{\cal{L}}^{(2)}_m = {\cal{O}}\left(-E_1+\sum_{i=2}^{m+2}E_i\right) \,,
\end{equation}
provided $m\le N-2$.

\subsection{Case 3: $\Sigma=T^2$, $n_{0+-}=n_{0-+}=1$}

This case arises when $V_2$ and $V_{\text{inst}}$ satisfy
\begin{equation}\left.\left(V_2\otimes V_{\text{inst}}^{-1}\right)\right|_{\Sigma}={\cal{O}} \,.
\end{equation}
We can rewrite this in terms of ${\cal{L}}$ as
\begin{equation}\text{deg}\left({\cal{L}}|_{\Sigma}\right)=-\text{deg}\left(\left[V_1\otimes V_2^{-1}\right]|_{\Sigma}\right) \,.
\end{equation}
Note that the object $\text{deg}([V_1\otimes V_2^{-1}]|_{\Sigma})$ is precisely what determines the spectrum of chiral massless fields localized on $\Sigma$.

To obtain a nontrivial superpotential contribution, we must also have
\begin{equation}n_{+0-}=n_{-0+}=1 \,.\label{c3cond}\end{equation}
Expressed in terms of the parameters in our ansatz \eqref{Lansatz}, these equations take the form
\begin{equation}3b_0+\sum_{i=1}^N b_i=0 \qquad \hbox{and} \qquad  b_0^2=\sum_{i=1}^N b_i^2-4 \,.
\end{equation}
No solutions to the second equation with $b_0\ne 0$ satisfy \eqref{b0cond} so we conclude that the only supersymmetric bundles ${\cal{L}}$ which satisfy \eqref{c3cond} have $b_0=0$.  In that case, the above equations take the simple form
\begin{equation}\sum_{i=1}^N b_i=0  \qquad \hbox{and} \qquad  \sum_{i=1}^N b_i^2 = 4 \,.
\end{equation}
Supersymmetric solutions to this equation are easy to enumerate and take the form
\begin{equation}{\cal{L}}^{(3)}={\cal{O}}\left(E_i+E_j-E_k-E_{\ell}\right)\qquad i,j,k,\ell \text{ distinct, }\end{equation}
provided $N\ge 4$.

\newpage

\bibliographystyle{JHEP}
\renewcommand{\refname}{Bibliography}
\addcontentsline{toc}{section}{Bibliography}

\providecommand{\href}[2]{#2}\begingroup\raggedright

\endgroup


\end{document}